\documentclass[aps,prd,twocolumn]{revtex4} 

\usepackage{graphicx}
\usepackage{epsfig}
\usepackage{psfig}
\usepackage{bm}% bold math
\def\gtap{\mathrel{ \rlap{\raise 0.511ex \hbox{$>$}}{\lower 0.511ex
   \hbox{$\sim$}}}} \def\ltap{\mathrel{ \rlap{\raise 0.511ex
   \hbox{$<$}}{\lower 0.511ex \hbox{$\sim$}}}} 
\newcommand{\beq}{\begin{equation}}

\newcommand{\eeq}{\end{equation}}
\newcommand{\bea}{\begin{eqnarray}}
\newcommand{\eea}{\end{eqnarray}}

\newcommand{\eV}{\mbox{$ \ \mathrm{eV}$}}

\newcommand{\MeV}{\mbox{$ \ \mathrm{MeV}$}}
\newcommand{\probm}{\mbox{$ \ \langle P_m \rangle$}}

\begin{document}
%@twocolumnfalse\endcsname 

\preprint{UCLA/03/TEP/20} 

\title{Pulsar kicks from a dark-matter sterile neutrino}

\author{George M. Fuller$^{1,5}$, Alexander Kusenko$^{2,3,5}$, 
Irina Mocioiu$^{4,5}$, and Silvia Pascoli$^{2,5}$ } 
\affiliation{$^1$Department of Physics, UCSD, La Jolla, CA 92093-0319\\
\mbox{$^2$Department of Physics and Astronomy, UCLA, Los Angeles, CA
90095-1547} \\
\mbox{$^3$RIKEN BNL Research Center, Brookhaven National
Laboratory, Upton, NY 11973} \\
\mbox{$^4$Department of Physics, University of Arizona, Tucson, AZ85721 }\\
$^5$KITP, UCSB, Santa Barbara, CA 93106 \\
}

%\date{July, 2003}

\begin{abstract}

We show that a sterile neutrino with mass in the 1--20 keV range and a small
mixing with the electron neutrino can simultaneously explain the origin of
the pulsar motions and the dark matter in the universe.  An asymmetric
neutrino emission from a hot nascent neutron star can be the explanation of
the observed pulsar velocities.  In addition to the pulsar kick mechanism
based on resonant neutrino transitions, we point out a new possibility: an
asymmetric off-resonant emission of sterile neutrinos.  The two cases
correspond to different values of the masses and mixing angles.  In both
cases we identify the ranges of parameters consistent with the pulsar kick,
as well as cosmological constraints.

\end{abstract}


%\vskip2.0pc]

\maketitle

Pulsars are known to have large velocities ranging from 100 to
1600~km~s$^{-1}$~\cite{astro}.  In contrast, the average velocity of an
ordinary star in our galaxy is 30~km~s$^{-1}$.  Born in a supernova
explosion of an ordinary star, the pulsar must, therefore, receive a
substantial ``kick'' at birth. The high angular velocities of pulsars may
also be explained by the same kick~\cite{sp}.  The origin of the pulsars
kicks remains an intriguing outstanding puzzle.

According to numerical simulations of supernova
explosions~\cite{explosion}, the asymmetries in the core collapse could not
account for a kick velocity of more than 300-600~km~s$^{-1}$.
Although the average pulsar velocity is in this range~\cite{astro}, 
there is a substantial population of pulsars with
velocities in excess of 700~km/s, including some with speeds over
1000~km~s$^{-1}$.  These velocities appear to be too large to arise from an
asymmetry in convection.   Evolution of close binary systems has
also been considered as a source of a kick velocity~\cite{b}.
Alternatively, it was argued~\cite{ht} that the pulsar may be accelerated
during the first few months after the supernova explosion by its
electromagnetic radiation, the asymmetry resulting from the magnetic dipole
moment being inclined to the rotation axis and offset from the center of
the star.  Both of these mechanisms, however, have difficulties explaining
the observed magnitudes of the pulsar kick velocities.

In this paper we point out a new way in which the pulsar kicks could be
generated from off-resonant active-sterile neutrino conversion in the
neutron star core.  In addition, we identify the parameters for which the
resonant conversions can take place.  Both cases are consistent with the
sterile neutrino being the  dark matter.   

The neutrinos emitted from a nascent hot neutron star carry a total energy
of 10$^{53}$~erg.  Only a 1\% asymmetry in the distribution of these
neutrinos could produce a kick consistent with the observed magnitudes of
pulsar velocities.  
Despite negligible magnetic moments, neutrinos are affected by the strong
magnetic field of the neutron star through their interactions with the
polarized electrons in the medium.  In particular, resonant neutrino
oscillations occur at different densities depending on the direction of the
magnetic field relative to the neutrino
momentum~\cite{polar1,polar3}.  Oscillations of active neutrinos
could explain the observed velocities of pulsars if the resonant conversion
$\nu_{\mu,\tau} \leftrightarrow \nu_e$ took place between two different
neutrinospheres~\cite{ks96,comment,ks98,barkovich,three}.
However, the neutrino
masses required for the resonant transition between the two
neutrinospheres, at density $10^{11}-10^{12} \, {\rm g \, cm}^{-3}$, are
inconsistent with the present limits on the masses of standard electroweak
neutrinos.

These limits do not apply, however, to sterile neutrinos that may have only
a small mixing with the ordinary neutrinos.  Resonant oscillations of
active neutrinos into sterile neutrinos have been proposed as a possible
explanation of the pulsar kicks~\cite{ks97}.  It was recently pointed out
that conversions of trapped neutrinos in the core of a hot neutron star can
make the effective mixing angle close to that in vacuum~\cite{Fuller}.
This has important implications for the pulsar kick mechanism because, in
the absence of matter suppression of the mixing, the non-resonant
production of sterile neutrinos can dominate.  This is asymmetric, and it
can explain the pulsar velocities as well.  The same sterile neutrino in
the keV mass range can be the dark matter in the universe~\cite{Fuller,dw}.
Sterile neutrino with the requisite masses and mixing appear in a number of
theoretical models~\cite{dolgov_review}, for example, in models with broken
mirror parity~\cite{mirror}.

We will assume that neutrinos have negligible magnetic moments and that they
have only the standard interactions with matter.  Unusually large magnetic
moments of neutrinos~\cite{voloshin}, as well as other kinds of new
physics~\cite{exotics} have also been considered as the origin of the
pulsar kicks.  We will further assume that only $\nu_e$ has a significant
mixing with $\nu_s$, characterized by $\sin^2 \theta \sim 10^{-11} -
10^{-7}$, while the oscillations between $\nu_{\mu,\tau}$ and $\nu_s$ are
suppressed by small mixing angle and/or matter effects.
 
For a sufficiently small mixing angle between $\nu_e$ 
and $\nu_s$, only one
of the two mass eigenstates,  $\nu_1$, is trapped.  The
orthogonal state, 
$| \nu_2 \rangle = \cos \theta_m | \nu_s \rangle + \sin \theta_m | \nu_e
\rangle , $
escapes from the star freely.  This state is produced in the same basic
urca reactions ($\nu_e+n\rightleftharpoons p+e^-$ and
$\bar\nu_e+p\rightleftharpoons n+e^+$) with the effective Lagrangian
coupling equal the weak coupling times $\sin \theta_m$.

Urca processes are affected by the magnetic field, and so the active
neutrinos are produced asymmetrically depending on the direction of their
momenta relative to the magnetic field.  Chugai~\cite{chugai} and Dorofeev
{\em et al.}~\cite{drt} have proposed that this asymmetry might explain the
pulsar kick velocities.  However, the asymmetry in the {\em production}
amplitudes does not lead to an appreciable asymmetry in the {\em emission}
of neutrinos because this asymmetry is washed out by scattering~\cite{eq}.
Therefore, the mechanism considered in Refs.~\cite{chugai,drt} does not
work.

However, since $| \nu_2 \rangle$ is not trapped anywhere in the star, the
asymmetry in its emission is exactly equal to the production asymmetry.
Depending on the parameters, this asymmetry can be as large as 25\%, just
as in the case of the active neutrinos~\cite{drt}. If the sterile neutrinos
carry away about 10\% of the thermal energy, the resulting overall
asymmetry can reach the required few percent, which is what one needs 
to explain the pulsar kick velocities.  

The production cross section of $\nu_2$ depends on the effective mixing
angle in matter $\theta_m$, which, in general, is different from the vacuum
mixing angle $\theta$:
\begin{equation}
\sin^2 2 \theta_m = 
\frac{(\Delta m^2 / 2p)^2 \sin^2 2 \theta}{(\Delta m^2 / 2p)^2 \sin^2 
2 \theta + ( \Delta m^2 / 2p \cos 2 \theta - V_m)^2}, 
\label{sin2theta}
\end{equation}
where the matter potential $V_m$ is positive (negative) for $\nu
(\bar{\nu})$, respectively; $p$ is the momentum.  For the case of $\nu_e$, 
which is relevant for our discussion,
$V_m$ reads:
\begin{equation}
V_m= \frac{G_{\!\!_F} \rho}{\sqrt{2} m_n} (3 Y_e-1+4 Y_{\nu_e}+2Y_{\nu_\mu}
+2Y_{\nu_\tau}). 
\end{equation}
In a core collapse supernova, the initial value of this matter potential is 
$V_m \simeq (-0.2 ...+ 0.5) V_0$, where $V_0= G_{\!\!_F} \rho / \sqrt{2}
m_n \simeq 3.8 \eV (\rho / 10^{14} \mathrm{g cm^{-3}}) $.  
The average probability of $\nu_e \rightarrow \nu_s$ conversion in  
presence of matter is  
\begin{equation}
\probm = \frac{1}{2} \left [ 
1+\left (\frac{\lambda_{\rm osc}}{2\lambda_{\rm s}} \right)^2
\right ]^{-1} \sin^2 2 \theta_m, 
\end{equation}
where we have included a possible suppression due to quantum
damping~\cite{dolgov}, which depends on the oscillation length
$\lambda_{\rm osc}$ and scattering length $\lambda_{\rm s}$.  

    It was pointed out in Ref.~\cite{Fuller} that, in the presence of sterile
neutrinos, rapid conversions can take place between different neutrino
flavors.  This, in turn, can drive the effective potential to its stable
equilibrium fixed point 
%
%\begin{equation}
$V_m \rightarrow 0.  $ 
%\end{equation} 
%
This equilibration takes place on a very short time scale and results in
the destruction of the initial asymmetry of $\nu_e$ over $\bar{\nu}_e$. (The
initial electron lepton number asymmetry in the supernova core depends on the
details of the prior stellar collapse history and, especially, the 
electron capture on heavy nuclei and neutrino transport in the in-fall
epoch.)  This, in turn, leaves the system out of  $\beta$-equilibrium.  On a
time scale $\sim$ms, electron capture reactions in the core will return
the system to beta equilibrium and the $\nu_e$ so produced will be converted. 
These processes will continue until both beta equilibrium and a steady
state equilibration are achieved. 
Once the equilibrium is achieved, the
effective mixing angle in matter is close to that in vacuum.  In
Ref.~\cite{ks97}, this effect was not taken into account, and only the
resonant emission from a thin shell was considered.  A larger mixing angle
makes a big difference in that the emission off resonance becomes 
possible from the entire volume of the neutron star core.

The equilibration mechanism relies on the fact that, in general, the
probability of $\nu_e \rightarrow \nu_s$ oscillations is different for
neutrinos and antineutrinos due to the opposite sign of $V_m$ in
eq.~(\ref{sin2theta})\footnote{ We note in passing that, for some extreme
values of magnetic fields, the $\nu_e$ and $\bar{\nu}_e$ oscillations into
$\nu_s$ may be driven by the magnetic field simultaneously~\cite{polar3}.}. 
The resulting change in the amounts of $\nu_e$ and $\bar{\nu}_e$ drives
$V_m$ to zero. Following Ref.~\cite{Fuller}, we estimate the time scale
for this process:
\begin{eqnarray}
\tau_{_V} &\simeq & \frac{V_m^{(0)} m_n}{ \sqrt{2} G_{\!\!_F} \rho}
\Big( \int d \Pi 
\frac{\sigma_{\nu}^{\rm urca} }{e^{(\epsilon_\nu -
    \mu_\nu)/T} +1}  
 \langle P_m(\nu_e \rightarrow \nu_s)\rangle
- \nonumber \\
& &
\int d \Pi  \frac{\sigma_{\bar\nu}^{\rm urca}}{e^{(\epsilon_{\bar{\nu}} -
    \mu_{\bar{\nu}})/T} +1}  
 \langle P_m(\bar{\nu}_e \rightarrow \bar{\nu}_s) \rangle
 \Big)^{-1}, 
\label{timeeq}
\end{eqnarray}
where $d \Pi= (2 \pi^2)^{-1} \epsilon_\nu^2 \ d \epsilon_\nu$, and
$V_m^{(0)}$ is the initial value of the matter potential $V_m$. 

Since the neutrino emission depends on the value of \probm, the time scales
for resonant and off-resonant conversions differ.  In the vacuum case,
$\Delta m^2 / (2 \langle E \rangle ) > | V_m|$, this mechanism is not
relevant, because $V_m$ is negligible from the start.  In the case of
resonance, $\Delta m^2 / (2 \langle E \rangle )\sim V_m $, the
conversions of $\nu_e$'s are strongly enhanced over the
$\bar{\nu}_e$'s.  Therefore the equilibration mechanism is very
efficient.  Using eq.~(\ref{timeeq}), we estimate the time:  
\begin{eqnarray}
\label{timeeqonres}
\tau_{_V}^\mathrm{on-res} & \simeq & \frac{2^5 \sqrt{2} \pi^2
  m_n}{G_{\!\!_F}^3  \rho}
\frac{ (V_m^{(0)})^6}{(\Delta m^2)^5 \sin 2 \theta } \left( e^{\frac{\Delta
  m^2 / 2 V_m^{(0)} - \mu}{T}} + 1 \right) \nonumber \\
& \sim  & \frac{2\times 10^{-9} s}{\sin 2 \theta} %\left(
\frac{ 10^{14}
%  \mathrm{g \, cm^{-3}}}{\rho} \right) 
  \frac{g}{ cm^{3}}} {\rho} %\right ) 
 \left(\frac{20 \, \mathrm{MeV}}{T} \right)^6 %\left( 
\frac{ \Delta m^2}{10 \,
   \mathrm{keV}^2} . %\right) 
\label{timeeqonresnumerical}
\end{eqnarray}
This holds as long as the resonance condition $V_m^{(0)} \simeq \Delta m^2
/(2 \langle E \rangle ) $ is satisfied.  The back-reaction effects on $V_m$ 
due to the change in $Y_{\nu_e}$ may slow down the equilibration.  Their
detailed analysis is beyond the scope of this paper; we do not expect the
back-reaction to alter the time scales significantly.

\begin{figure}
\epsfxsize=8cm 
\epsfbox{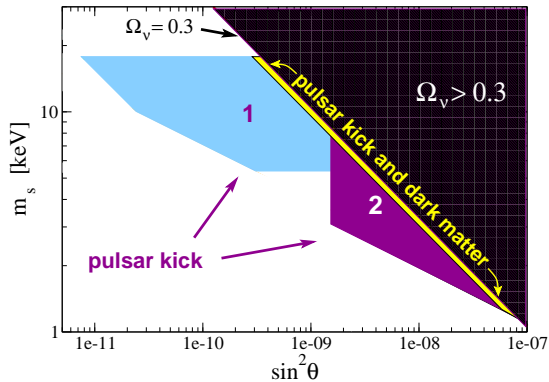}
\caption{ Neutrino conversions can explain the pulsar kicks for values of
parameters either in region 1~(Ref.~\cite{ks97}),  
or in region 2 [present work]. In a band near the {$\Omega_\nu = 0.3$}
line, the sterile neutrino can be 
the dark matter.}
\label{figure:parameters}
\end{figure}

If neutrinos are converted off-resonance, it is necessary to evaluate the
integral in eq.~(\ref{timeeq}) taking into account that both the chemical
potential and the reaction rates for neutrinos and antineutrinos are
different.  An approximate evaluation of the time scale yields: 
\begin{eqnarray}
\label{timeeqoffres}
 \tau_{_V}^\mathrm{off-res}  & \simeq &
\frac{4 \sqrt{2} \pi^2 m_n}{G_{\!\!_F}^3 \rho}
\frac{ (V_m^{(0)})^3}{(\Delta m^2)^2 \sin^2 2 \theta } \frac{1}{\mu^3}
\\
& \sim & 
 \frac{6 \times 10^{-9} s}{\sin^2 2 \theta} 
\left (\frac{V_m^{(0)}}{0.1 \mathrm{eV}} \right )^3 
\left (\frac{50 \mathrm{MeV}}{\mu} \right )^3 \left ( \frac{ 
10 \mathrm{keV}^2
}{\Delta m^2
} \right )^2. \nonumber   
\label{timeeqoffresnumerical}
\end{eqnarray}
Requiring that the off-resonance equilibration takes place 
during the first second, 
imposes a combined lower bound on $\Delta m^2$ 
and  $\sin^2 2\theta$.  
In  Fig.~\ref{figure:parameters}
this bound, for $V_m^{(0)}=0.1 \, \eV$, $\mu=50$~MeV, $T=20$~MeV,  
limits region 2 from below.
($T=20 \MeV$ is  plausible for the immediate post core bounce
conditions, where models suggest a core temperature between tens of MeV and
about $60\,{\rm MeV}$~\cite{prakash}.)

The probability of oscillation depends on the values of $ \Delta m^2$,
$\sin \theta$, and $V_m$.  One can identify three different regimes. 
(i)~the ``vacuum'' case: $\Delta m^2 / (2 \langle E \rangle ) \gtap |V_m|$;    
(ii)~the resonance regime, $\Delta m^2 / (2 \langle E \rangle ) \approx
V_m$, in which the conversion probability is strongly enhanced;   
(iii)~the ``matter-suppressed'' oscillations, for
$\Delta m^2 / (2  \langle E \rangle ) \ll   V_m$, which have a vanishing
probability. 
%
%Finally, (iv) if the ``equilibration'' of $V_m \rightarrow 0$ occurs, 
%the matter potential starts out non-zero and is driven to zero by
%oscillations. 

%%%%%%%%%%%%%%%
%%%%%%%%%%%%%%%%%%5

Whether or not a resonance occurs anywhere in the core depends on the mass
of the sterile neutrino and its mixing angle with $\nu_e$.  The resonance
occurs at some point in the core for parameters in the range marked
``1'' in Fig.~\ref{figure:parameters}.  The lower boundary of this region
comes from enforcing the adiabaticity and the weak damping conditions.  The
vertical span reflects the range of densities in the outer core of a
neutron star. 

{\bf The case of resonant emission.}  If the resonance condition is
satisfied (region 1 in Fig.~\ref{figure:parameters}), the emissivity is
strongly enhanced in the region of resonant density, while it is suppressed
elsewhere.  The resonant region is a thin layer, whose deviation from a
spherical shape is due to the electron polarization in the magnetic
field~\cite{polar1,polar3,ks96,ks97}. For small mixing angles (region 1),
the time $\tau_{_V}^\mathrm{off-res}$ is longer than a few seconds for some
reasonable initial conditions.  Hence, the matter potential is not affected
outside the resonant region. Therefore, in region 1 the pulsar kick
mechanism works as described in Ref.~\cite{ks97}, where the equilibration
was not considered~\footnote{
Refs.~\cite{ks96,ks98} and~\cite{barkovich} are in agreement regarding the
pulsar kick mechanism based on the active neutrino conversions. The
approach used in Ref.~\cite{barkovich} can be modified to apply to the
case of sterile neutrinos~\cite{ks97}.}.

{\bf Off-resonance emission after $\bm {V_m}$ has equilibrated to zero.} To
get the pulsar kick, we require that a substantial fraction of energy be 
emitted in sterile neutrinos.  The dominant process of neutrino production
in the core is urca. One can estimate the emissivity in sterile neutrinos
following Ref.~\cite{Fuller}:
\begin{equation}
{\cal E} \sim \int d \Pi \, \epsilon_\nu
\frac{\sigma_{\rm{urca}} }{ 
e^{(\epsilon_\nu - \mu)/T} +1 } \ 
\frac{1}{2}
\langle P_m \rangle.  
\label{emissivity}
\end{equation}
The emissivity depends on the value of \probm.  In the ``vacuum'' regime
the emissivity is proportional to $\sin^2 2 \theta$ and is independent of
$\Delta m^2$.  The total energy emitted per unit mass is simply the
emissivity multiplied by the time of emission, which is 5--10 s.  For given
values of $\mu$ and $T$, the requirement of emitting a certain fraction of
the energy in sterile neutrinos translates into a lower bound on the
allowed values of $\sin^2 2 \theta$, as shown in
Fig.~\ref{figure:parameters}.  Such a bound is independent of $\Delta
m^2$, as long as the $V_m\simeq 0$ condition is satisfied.  If the
equilibration process is efficient, $V_m \approx 0$ and the matter effects
are negligible.  Then the oscillations $\nu_e \rightarrow \nu_s$ are
controlled by the vacuum mixing angle $\sin^2 2 \theta$.

\begin{figure}
\epsfxsize = 8cm  
\epsfbox{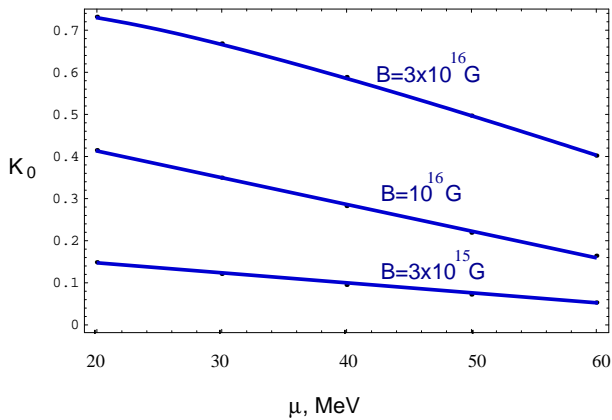}
\caption{The fraction of electrons in the lowest Landau level as a function
of $\mu$ for $T=20 \, \MeV$.  The value of the magnetic field in the core
of a neutron star is shown next to each curve.  }
\label{figure:asymmT20}
\end{figure}

As discussed by Dorofeev {\em et al.}~\cite{drt}, the urca processes result
in an asymmetric neutrino production if the electrons are polarized. This 
asymmetry arises from the spin-multiplicity factor.  Only the electrons in
the lowest Landau level ($n=0$) contribute to this asymmetry.  The number
of neutrinos $dN$ emitted into a solid angle $d\Omega $ can be written as
%
%\begin{equation}
$
\frac{dN}{d\Omega}= N_0(1+ \epsilon \cos \Theta_\nu ), 
$
%\end{equation} 
where $\Theta_\nu$ is the angle between the direction of the magnetic field
and the neutrino momentum, and $N_0$ is some normalization factor.  The
asymmetry parameter $\epsilon$ is equal
\begin{equation}
\epsilon = \frac{g_{_V}^2-g_{_A}^2}{g_{_V}^2+3g_{_A}^2} 
k_0   
\left ( \frac{\cal  E_{\rm s}}{\cal E_{\rm tot}} \right ),  
\end{equation} 
where $g_{_V}$ and $g_{_A}$ are the axial and vector couplings,
${\cal  E_{\rm tot}}$ and $ {\cal E_{\rm s}} $ are the total
neutrino luminosity and that in sterile neutrinos, respectively.
Following  Ref.~\cite{drt}, we obtain 
%
%%%
\begin{equation}
k_0 = \frac{F(0)}{F(0)+2 \sum_{n=1}^\infty F(n)}, 
\label{k1}
\end{equation}
where
\begin{equation}
F(n) = \int_{0}^{\infty} d p \frac{ \left (m_n - m_p - \sqrt{p^2 +
  (m_e^*)^{2}+2Bn}\right )^2}{
\exp \{ \frac{\sqrt{p^2 + (m_e^*)^2+2 B n} - \mu}{T} \} + 1  }
\label{F(n)}
\end{equation}
Here $n$ is the electron Landau level, and we have assumed that protons and
neutrons are non-relativistic, non-degenerate, and that their polarization
can be neglected.  The effective electron mass at high density, $m_e^*$,
remains below 10-11~MeV everywhere in the star~\cite{prakash,hardy}.  Its
exact value does not affect the expression for $F(n)$ in eq.(\ref{F(n)}) by
more than a few per cent. If only the lowest Landau level is populated, then
$k_0=1$.  The weaker the magnetic field, the more levels are populated, the
smaller is the fraction $k_0$.  In Fig.~\ref{figure:asymmT20} we show $k_0$
as a function of chemical potential $\mu$ for different values of the
magnetic field $B$.  We took the temperature to be $T=20 \MeV$.

The ratio $\left ( {\cal  E_{\rm s}} / {\cal E_{\rm tot}} \right )= 
 {\cal  E_{\rm s}} / ({\cal E_{\rm s}}+{\cal E_{\nu }} ) $ can
be estimated from the following: 
\begin{equation}
\left ( \frac{\cal  E_{\rm s}}{\cal E_{\nu}} \right ) \sim  
\sin^2 \theta \, 
\left ( \frac{T_{\rm core} }{T_{\nu-{\rm
      sphere}}} 
  \right )^6  \, f_{_{\rm M}}\,  f_{\rm d.o.f.},  
\end{equation}
where $f_{_{\rm M}}< 1$ is the fraction of enclosed mass of the core from
which the emission of sterile neutrinos is efficient, and $ f_{\rm d.o.f.}
\ge 1$ is an enhancement due to a possible increase in the effective
degrees of freedom at high density~\cite{prakash}.   
We take the core temperature $T_{\rm
core} \approx (20-50) {\rm MeV}$, while the neutrinosphere temperature is $
T_{\nu-{\rm sphere}}\approx (2-10) {\rm MeV}$.  Hence, for $\sin^2 \theta
\sim 10^{-8}$ the ratio $\left ( {\cal E_{\rm s}} / {\cal E_{\rm tot}}
\right )$ can be in the range
\begin{equation}
r_{\cal E}= 
\left ( \frac{\cal  E_{\rm s}}{\cal E_{\rm tot}} \right ) \sim  0.05 - 0.7.
\label{rE}
\end{equation}
A higher ratio is in conflict with the observation of neutrinos from 
SN1987A, while a lower value does not give a sufficient pulsar kick.  Our
point here is that, for reasonable values of parameters, the range in
eq.~(\ref{rE}) is {\em possible}, even if not necessary.
% 
%%%%%%%%%%%%%%%
%%%%%%%%%%%%%%%%%%

Finally, the magnitude of the asymmetry is 
\begin{equation}
\epsilon \sim 0.02 
\left ( \frac{k_0}{0.3}\right ) 
\left ( \frac{ r_{\cal E}}{0.5} \right), 
\end{equation}  
which can be of the order of the requisite few per cent for magnetic fields
$10^{15}-10^{16}$~G, as can be seen from Fig.~\ref{figure:asymmT20}.  

The magnetic field at the surface of an average pulsar is estimated to be
of the order of $10^{12}-10^{13}$G \cite{pulsar_review}.  However, the
magnetic field inside a neutron star may be as high as $10^{16}$G
\cite{pulsar_review,dt,r,magnetars}.  The existence of such a strong
magnetic field is suggested by the dynamics of formation of the neutron
stars, as well as by the stability of the poloidal magnetic field outside
the pulsar~\cite{dt}.  Moreover, the discovery of soft gamma repeaters and
their identification as magnetars~\cite{magnetars}, {\em i.e.}, neutron
stars with {\em surface} magnetic fields as large as $10^{15}$~G, gives one
a strong reason to believe that the interiors of many neutron stars may
have magnetic fields as large as $10^{15}-10^{16}$~G and that only in some
cases this large magnetic field breaks out to the surface.  There is also a
plausible physical mechanism by means of which such a large magnetic field
can form inside a neutron star through the so called $\alpha - \Omega$
dynamo effect~\cite{zeldovich,dt}.  Nuclear matter inside a cooling neutron
star is expected to have a sufficient amount of convection to generate
magnetic fields as large as $10^{16}$G~\cite{dt}.

The ratio of luminosities $r_{\cal E}$ translates into an upper bound on
the mixing angle.  The allowed region for the masses and mixing angles is
marked ``2'' in Fig.~\ref{figure:parameters}.

Gravity waves from a pulsar kick due to neutrino conversions might be
observable~\cite{cuesta} and could provide a way to test this mechanism.

To summarize, we have analyzed the pulsar kick resulting from active to
sterile neutrino conversions.  We have shown that, in addition to resonant
transitions~\cite{ks97}, the off-resonant neutrino oscillations can also
result in a pulsar kick consistent with observations.  The two mechanisms
require different masses and mixing angles.  We have identified the range of
parameters for which each of the two mechanisms can explain
the observed velocities of pulsars.  Part of this range is consistent with
the sterile neutrino being the dark matter with $\Omega_\nu^{^\mathrm{WDM}}
\approx 0.27$.  

A.K. thanks M.~Prakash for discussions. The work of G.F. was supported in
part by the NSF grant PHY-00-99499; that of A.K. and S.P. was supported in
part by the DOE Grant DE-FG03-91ER40662 and the NASA grant ATP02-0000-0151.
I.M. was supported by the DOE Grant DE-FG02-95ER40906.  The authors thank
KITP at UCSB and CERN (S.P.) for hospitality.

\end{document}